\documentclass[pdflatex,sn-mathphys-num]{sn-jnl}

\usepackage{graphicx}%
\usepackage{multirow}%
\usepackage{amsmath,amssymb,amsfonts}%
\usepackage{amsthm}%
\usepackage{mathrsfs}%
\usepackage[title]{appendix}%
\usepackage{xcolor}%
\usepackage{textcomp}%
\usepackage{manyfoot}%
\usepackage{booktabs}%
\usepackage{algorithm}%
\usepackage{algorithmicx}%
\usepackage{algpseudocode}%
\usepackage{listings}%
\usepackage{subcaption}

\theoremstyle{thmstyleone}%
\theoremstyle{thmstyletwo}%

\theoremstyle{thmstylethree}%

\begin{document}

\title[Article Title]{PolicyStory: Leveraging Large Language Models to Generate Comprehensible Summaries of Policy-News in India}


 \author*[1]{\fnm{Aatif Nisar} \sur{Dar}}\email{aatif.dar11@gmail.com}

\author[2]{\fnm{Aditya Raj} \sur{Singh}}\email{adityarajsingh.8502@gmail.com}

 \author[2]{\fnm{Anirban} \sur{Sen}}\email{anirban.sen@ashoka.edu.in}

\affil*[1]{\orgdiv{Department of Computer Science}, \orgname{Indian Institute of Technology Delhi}, 
\state{New Delhi}, \country{India}}

\affil[2]{\orgdiv{Department of Computer Science}, \orgname{Ashoka University}, 
\city{Sonipat}, 
\state{Haryana}, \country{India}}



\abstract{In the era of information overload, traditional news consumption through both online and print media often fails to provide a structured and longitudinal understanding of complex socio-political issues. To address this gap, we present \textit{PolicyStory}, an information tool designed to offer lucid, chronological, and summarized insights into Indian policy issues\footnote{The phrases \textit{policy event} and \textit{policy issue} are used interchangeably in the paper, and they indicate any event/discussion around policymaking}. PolicyStory collects news articles from diverse sources, clusters them by topic, and generates three levels of summaries from longitudinal media discourse on policies, leveraging open source large language models. A user study around the tool indicated that PolicyStory effectively aided users in grasping policy developments over time, with positive feedback highlighting its usability and clarity of summaries. By providing users a birds' eye view of complex policy topics, PolicyStory serves as a valuable tool for fostering informed civic engagement.}


\maketitle

\section{Introduction}\label{sec1}

In the past, newspapers or print media served as the primary source of social and political information. Readers relied heavily on the textual content provided in print media to stay informed about social and political developments relevant to their interests. The structure of newspaper articles reflect varying levels of information granularity. Headlines function as concise summaries, facilitating quick browsing and efficient information retrieval. The body of the text, often accompanied by images, offers more comprehensive and nuanced insights into the events. Depending on individual preferences, readers can either skim multiple articles by reading only the headlines, or engage more deeply with selected articles for detailed understanding.

With the emergence of new modes and systems of information dissemination, users increasingly encounter the challenge of information overload. The vast and continuously growing volume of content distributed across diverse multimodal platforms makes it practically impossible for individuals to consume all available information on important issues comprehensively. Moreover, the proliferation of sources, varying widely in credibility and quality, further complicates the process of discerning accurate and trustworthy information. This not only hampers informed decision-making but also contributes to the spread of misinformation and confusion in the public sphere. Finally, users often lack even a high-level overview of critical developments, such as those related to policymaking, owing to the sheer complexity and drabness of information presented.

The growing dominance of short-form video content, often referred to as \textit{reel culture}, has raised concerns about its effects on users' attention spans and their ability to engage with in-depth, high-quality news content~\cite{bib6,bib7}. While not all readers can be expected to understand each policy event in depth, it is essential for everyone to carry a high level overview of them. Additionally, for policy issues that are longitudinal or periodic in nature (e.g., Central/State Elections and Union Budget) users must be aware of the evolution of common policy related topics. For example, in the context of the Union Budget, adopting a high-level overview of the trends in defense expenditure—encompassing growth/decline over time—can provide critical insights into the government’s fiscal priorities and strategic orientations. This is especially important since knowledge around policy events aids users in participating meaningfully and responsibly in the democratic processes. This in turn leads to developmental outcomes that are beneficial to the society at large.

As a first step towards providing a bird’s eye view of policy issues to users, we have developed an information tool named \textit{PolicyStory} that enables users in getting summarized information around Indian policy issues. PolicyStory leverages open source Large Language Models (LLMs) like Llama (Llama-3.2-1B) to summarize and present policy related news summary in a chronological (yearly) fashion. The primary advantage of this tool, and its difference from other systems of similar nature, is that: 
\begin{enumerate}
    \item Alongside summarizing policy specific news for users, it also presents the summaries in a chronological manner over a period of time, for recurring or long term policy related events.
    \item It extracts the salient topics within the policy issues that are in discussion and development over a certain period of time. The summaries are generated for each topic separately. 
    \item It provides a user the option to peruse summaries at three levels – a brief one-paragraph summary (also called the L1 summary), a detailed summary (L2 summary) for in depth understanding of the event, and a numeric summary providing quantitative data around the event. The L2 summary is presented in the form of storytelling, while retaining all of the technical details. This functionality allows users to access and browse summaries in accordance with their interest and bandwidth.
    \item The tool also defines policy related jargon for the users in simpler terms for easier understanding.
\end{enumerate}
The current paper discusses the development and features of the first pilot of PolicyStory, which presents information around two national policy events in India – the Union Budget (2019-2014), and the Farmer’s Protests (2021-2024). \textit{The Union Budget} is the annual financial statement of the estimated receipts and expenditure of the government for a financial year. This statement serves as a financial blueprint for the upcoming fiscal year, forecasting economic conditions and aligning the Government of India's spending with its policy objectives. The budget is discussed in the Indian Parliament on the 1st of February each year, so as to allow its adoption in the upcoming fiscal year that starts in April. The \textit{Indian Farmers' Protest} primarily refers to a series of protests against three farm acts that were passed by the Parliament of India in September 2020. Notably, the Farmers' Protests occurred in two distinct phases. The first began in August 2020 and lasted till December 2021, culminating in the repeal of the contentious farm laws~\cite{bib2}. A second wave of protests emerged in early 2024, with renewed demands around Minimum Support Prices (MSP) and farmer welfare.

A high level architecture diagram of the tool is shown in Figure \ref{fig3}. PolicyStory presents relevant policy information (or summaries) to the users through the following steps: (A) Collecting news articles using a keyword based approach from \textit{Media Cloud}~\cite{bib11} (a large scale repository of news from various sources), (B) Extracting topics of discussion from the news articles collected, (C) Generating L1 and L2 summaries of articles for each topic, (D) Generating numeric summaries (key-value pairs) around commonly discussed points over time, and (E) Presenting these summaries in a comprehensible form to users through a web based interface.

The first pilot of the tool \cite{PolicyStory} was deployed in the public domain\footnote{URL: https://neural-times.vercel.app/} and a user study was carried out, to obtain feedback on the tool and its usability. The user study revealed that PolicyStory generally helped users in getting an easy-to-understand, longitudinal summary of policy events. Several participants concurred around the difficulty of obtaining similar chronological information around policy issues through news articles directly. The feedback around usability and features of the tool included comments on interface smoothness, ease of navigation, clarity and impact of summaries, flexibility in summary depth, visibility of numeric information, and the importance of citing credible sources to enhance trust. Overall, from the initial user study, it was evident that PolicyStory was indeed an important step in the direction of providing comprehensible view of policy issues of importance.

As part of future work, we intend to enhance the system by incorporating a broader range of policy issues that hold national and regional significance. In addition, while the current collection of events is static, we aim to provide users with the capability to dynamically query the system for specific policy issues of interest using relevant keywords. Our user study also requires expansion to include feedback from a more diverse and larger participant base, thereby increasing the generalizability of our findings.

Furthermore, connecting the summaries with appropriate source citations and expert opinions can be an important feature of the tool, enhancing its credibility. It will also be valuable to examine and present the impact of these policy discussions in a comprehensive fashion, by augmenting the system with analysis of government documents around policy-making. Finally, within each policy issue, it will be beneficial to offer users real-time analytical visualizations dynamically, based on queried keywords (e.g.  \textit{capital expenditure}) with quantitative components. Ongoing research will focus on integrating these additional axes of analysis and features into PolicyStory.

\section{Related Work}
This section describes the previous literature we referred to for the study, along three axes of analysis. First, we discuss work done around readability of text data. Second, we touch upon studies that propose systems to enable balanced news consumption. Finally, we discuss some attempts at text summarization using LLMs.

\subsection{Systems to Deliver Balanced News}
Several prior studies have focused on building systems for news aggregation, aiming either at balanced consumption or coherent delivery. Park et al.~\cite{park2009newscube} introduced \textit{NewsCube}, a system that detected event-related topics and ensured viewpoint plurality, allowing users to adjust content preferences. Munson et al.~\cite{munson2013encouraging} developed a browser widget that nudged users toward politically balanced news by analyzing their reading behavior. Park et al.~\cite{park2010aspect} proposed a framework for classifying news articles by perspective, enabling users to explore multiple viewpoints and form their own understanding.

Our work is inspired by these efforts. Although we do not focus directly on balance or fairness, our system offers a bird’s eye view of policy issues by summarizing news across a wide range of topics. Presenting summaries along multiple topical axes ensures users gain exposure to different facets of policy discourse. While several Indian news aggregators exist~\cite{reddy2024smart}, they typically emphasize current affairs. In contrast, \textit{PolicyStory} delivers topic-based, chronological, and longitudinal summaries, enabling users to track the evolution of complex policy issues over time rather than just consuming cross-sectional snapshots.

\subsection{Text Summarization using LLMs}

Large Language Models (LLMs) have made Automatic Text Summarization (ATS) highly effective. Pre-trained on diverse datasets, LLMs excel in few-shot and zero-shot learning, enabling high-quality summaries with minimal input. Unlike earlier supervised models, LLMs combine both \textit{abstractive} and \textit{extractive} methods, leading to more flexible and superior summaries.

LLMs have been applied across various domains~\cite{chuang2024spec,hartl2022applying,zogan2021depressionnet}, overcoming the limitations of earlier summarization methods. Recent approaches often adopt hybrid models that blend extractive and abstractive techniques~\cite{ghadimi2022hybrid}. Several studies have focused on advancing summarization: Ding et al.~\cite{10685184} used GPT-4o for question-driven personalized summaries; Fang et al.~\cite{fang2024multi} proposed a multi-LLM framework outperforming single-LLM methods; Shah et al.~\cite{shah2025topic} employed topic-based summarization for large unstructured texts. In addition, Sahu et al.~\cite{sahu2024mixsumm} addressed summarization in low-resource languages using open-source LLMs. Song et al.~\cite{song2024finesure} introduced \textit{FineSurE}, a fine-grained summary evaluator assessing completeness, conciseness, and faithfulness, going beyond standard ROUGE metrics. Jiang et al.~\cite{jiang2024empirical} demonstrated strong results in long-text summarization using a combination of smaller language models and LLMs.

While many of these techniques are relevant to policy news summarization, our system focuses on delivering lucid, concise, and chronologically coherent summaries using zero-shot LLM-based methods. To mitigate potential information loss, we generate summaries at three levels: (i) a 30-second brief (L1), (ii) a detailed narrative (L2), and (iii) a numeric summary capturing quantitative insights. This multi-layered approach aims to support users in building a clear and structured understanding of complex policy developments over time.

\subsection{News Readability Analysis}
Text can express the same idea in varied ways using different linguistic features like sentence length, word choice, and structure. These variations affect comprehension based on user preferences. Prior research highlights linguistic patterns that improve readability~\cite{dalecki2009news, kleinnijenhuis1991newspaper}. Yet, news readability has declined over time, despite journalistic norms discouraging complex prose~\cite{danielson1992journalists}. The use of jargon and complex syntax further hampers understanding~\cite{Patoko2014TheIO, Khawaja2014MeasuringCL, fix2020effect}.
\par
Motivated by these findings, we propose an end-to-end system to deliver policy news in a comprehensible format. Focusing on document length and complexity, we provide layered summaries and explain policy-related jargon to enhance clarity. Our user study supports the effectiveness of these readability-focused features. Future work will integrate additional metrics to quantify readability.

\section{Methodology}
This section describes the methodology used to develop PolicyStory, which consists of four main components: Data Collection, Content Summarization, Numeric Summary Generation, and User Interface Development, which are detailed here. 
The overall workflow of the system is shown in Figure ~\ref{fig3}.

\subsection{Data Collection}
The first step of our work focused on gathering relevant news articles for two national policy issues in India: the Union Budget (2019–2024) and the Farmers' Protests (2020–2024). Articles were sourced for both of the events from \textit{Media Cloud}~\cite{bib1}, an open-source repository that aggregates news from a wide range of media outlets, including both national and regional news media. To ensure comprehensive and relevant retrieval of policy-related news content, we crafted targeted keyword queries for each policy issue. These queries were designed to capture the core terms and associated discourse commonly found in media coverage (Table \ref{tab1}).

    \begin{table}[h]
        \centering
        \begin{tabular}{|p{4cm}|p{8cm}|}
        \hline
         Event & Keywords  \\
         \hline
         Farmers' Protests &  \texttt{"farmers" AND ("protest" OR "agitation" OR "farm laws" OR "MSP" OR "march to Delhi")}\\
         \hline
         Union Budget & \texttt{"budget" AND ("finance minister" OR "union budget" OR "fiscal policy" OR "tax reforms" OR "Nirmala" OR "budget speech" OR "budget allocation" OR "fiscal deficit")}\\
         \hline
        \end{tabular}
        \caption{Keywords used for data collection from Media Cloud for each policy event}
        \label{tab1}
    \end{table}
    
This approach enabled us to extract a large and thematically rich set of articles for subsequent analysis. Due to limited computational resources, but with the aim of ensuring both temporal balance and comprehensive coverage, 2000 articles were collected per year for each policy issue, for the period of data collection. A stratified sampling strategy was employed to distribute article selection evenly across the months of each calendar year. This ensured that the dataset captured the evolution of discourse over time rather than being biased toward specific time periods with high media attention. 
\par
Since the Farmers' Protests began in August 2020, we collected data from that month onward to reflect the actual onset of media coverage. Although the protest laws were officially repealed in December 2021~\cite{bib2}, we chose to generate stories for the year 2022 to reflect the broader national discourse that continued beyond the repeal. In February 2024, Farmers’ Protest 2.0 began, driven by renewed demands for a legal guarantee on Minimum Support Prices (MSP) and concerns over unfulfilled promises from the previous agitation~\cite{bib3}. Hence we decided to generate stories for the year 2024 as well.
\par
The Union Budget in India follows an annual cycle. However in election years, the budget is typically presented twice, first as an interim budget before elections, and then again as a full budget by the incoming government. Our study collected data on Indian Union Budgets from 2019 to 2024, including both regular annual budgets and the two-part budgets (interim and full).
\par
As a next step, the URLs extracted from Media Cloud were parsed using the \textit{Newspaper3k}\footnote{Newspaper3k: Article scraping \& curation - https://newspaper.readthedocs.io/en/latest/} library, which returned structured metadata such as article titles, publication dates, author lists, and the main body of text. Media Cloud has most of its data available from after 2020. As a result, we were unable to obtain the full set of 2000 articles for the years 2019 and 2020. After parsing all available articles through the Newspaper3k API, limiting to 2000 articles per year, Table \ref{tab2} displays the number of articles from which we were able to successfully extract text from. 

\begin{table}[h]
\centering
\caption{Number of articles from which we were able to successfully extract text}
\begin{tabular}{lcccccc}
\toprule
Topic & 2019 & 2020 & 2021 & 2022 & 2023 & 2024 \\
\midrule
Union Budget (2019-2024) & 157 & 218 & 1,467 & 1,652 & 1,453 & 1,676 \\
Farmers' Protest (2020-2024) & -- & 1,456 & 1,508 & 1,667 & -- & 1,669 \\
\bottomrule
\end{tabular}
\label{tab2}
\end{table}

\subsection{Content Summarization}
The content summarization step focused on transforming the collected news articles into structured, readable stories that offer users both a quick overview and a deeper understanding of complex policy developments. To achieve this, we used the open-source LLM \texttt{Llama-3.2-1B}, implemented through the \textit{Hugging Face Transformers} library~\cite{bib4}. This open-source model was chosen because it produced clear and coherent narratives even when working with large volumes of unstructured news text~\cite{bib5}. Additionally, the free to use LLM was easy to run on the available institutional infrastructure (Google Colab, A100 GPU), which made this process both accessible and efficient.
\par
In order to be inclusive around the user bandwidth of policy related news consumption, we used a two-level storytelling strategy. Level 1 (L1) stories are short, one-paragraph overviews designed for quick reading. These stories give a quick overview of key developments while defining jargon and are generally readable within 30 seconds to 1 minute. Level 2 (L2) stories are longer and detailed summaries presented in a story-like fashion, capturing key developments over time while explaining any technical or policy-specific terms.
\par
To ensure that the stories made sense in terms of both the timeline of the event and topics (aspects) involved, we first grouped the news articles corresponding to each policy issue by date and theme. The date column was parsed using the Newspaper3k API, so we used that column to split the news articles according to date. We then categorized the articles into clusters or themes based on their content. Each article's title and first paragraph were used as inputs to a prompt-driven, LLM-based classifier, which assigned the article to the most appropriate topic cluster. For the Farmers' Protests dataset, 5 clusters were created. For the Union Budget dataset, 8 clusters were created. More details including the definition of the clusters can be found in the supplementary material \cite{supplementary}.
\par
\begin{figure}[h]
  \centering
  \includegraphics[width=\linewidth]{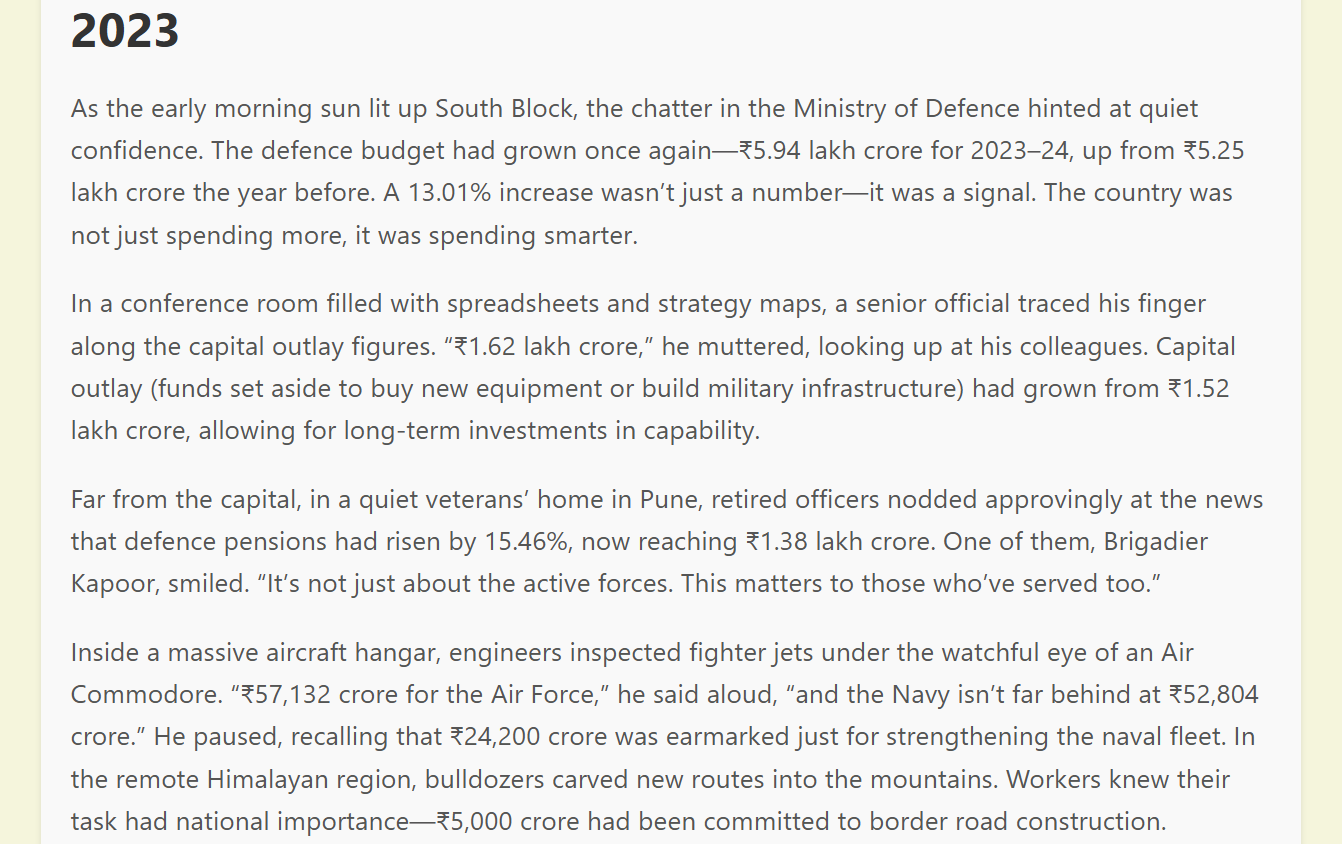}
  \caption{A snippet of an L2 story from the Defense theme of 2023.}
  \label{fig1}
 \end{figure}

In order to generate coherent and meaningful stories (L2 stories), we needed article content with most of the essential details preserved. We thus had two options: for each theme/topic, either feed the entire text of all articles or use long, detailed summaries consisting of 5–8 sentences\footnote{Article summary length of 5-8 sentences was experimentally decided.} per article. We opted for the latter approach to avoid exceeding the context window limit of 4,096 tokens of the Llama model. The LLM generated summaries were still comprehensive, containing all the essential information from each article when checked manually (20 articles from each topic were checked by the authors). To again avoid exceeding the context window limit, we processed the summaries of articles in batches of 20 for generating the final L2 stories. Figure \ref{fig1} contains a snippet of an L2 story from the Defense theme of 2023. This is only a portion of the complete narrative. The full story can be accessed on the website.


\begin{figure}[h]
  \centering
  \includegraphics[width=\linewidth]{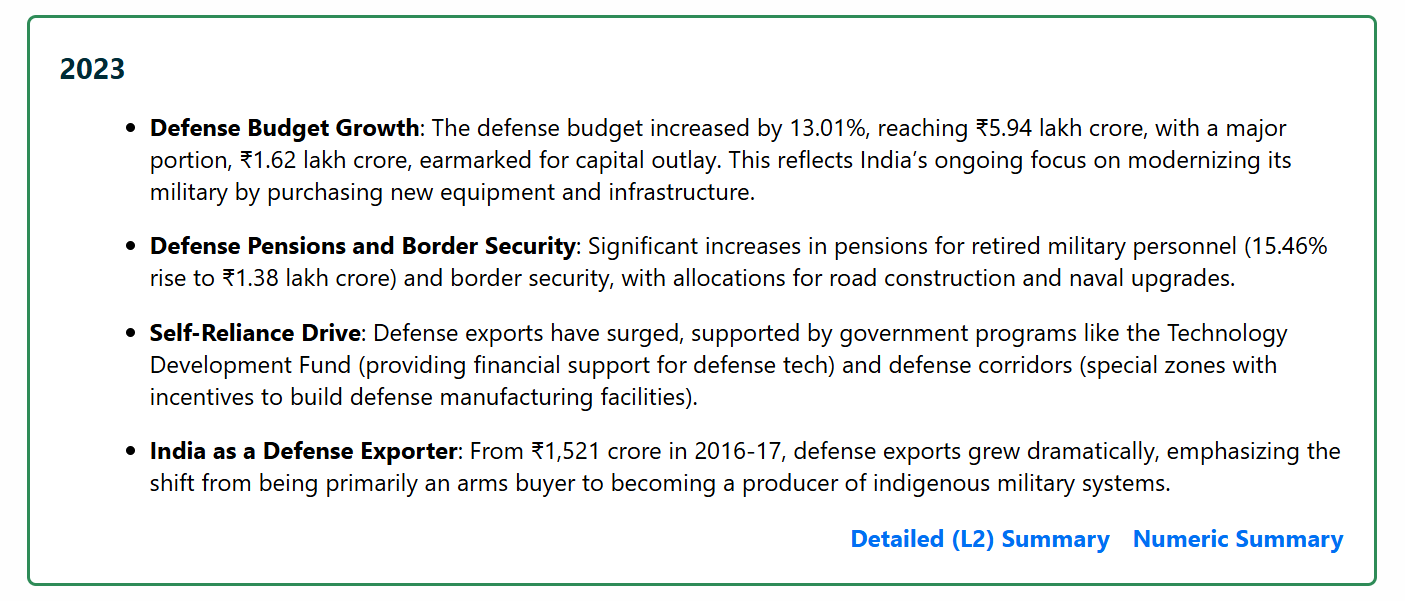}
  \caption{A snippet of an L1 story from the Defense theme of 2023.}
  \label{fig2}
 \end{figure}

\begin{figure}[h]
  \centering
  \includegraphics[width=\linewidth]{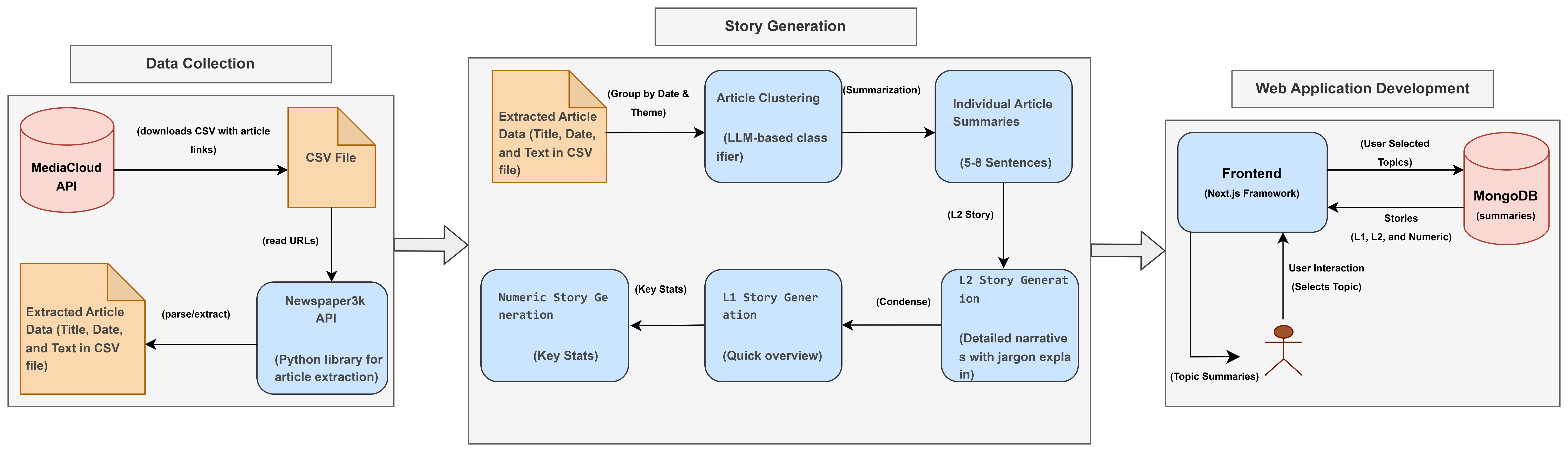}
  \caption{This figure illustrates the complete workflow of the PolicyStory system, highlighting each key component from data acquisition to user-facing delivery. The process starts by sourcing news articles from Media Cloud using keyword-based queries tailored to each policy topic. After parsing with Newspaper3k, articles are clustered by date and theme using an LLM-based classifier. This ensures that content is organized meaningfully before summarization.\\
  Each article is summarized individually, and then cluster-level summaries are generated to form detailed L2 stories and concise L1 summaries. Numeric data points are extracted from L2 stories to provide quick insights. All content is stored in a MongoDB database and served through a React + Next.js web application. Users can explore stories by year or topic, choose between L1, L2, or numeric views, and access policy-related jargon definitions directly in the interface. A separate section has been added to the supplementary material \cite{supplementary} detailing the development of the web application, including the tools and technologies used.}
  \label{fig3}
  
 \end{figure}

Once the L2 summaries were ready, we used them to create shorter L1 summaries, again for each topic and for each year. These were short summaries with clear overviews that highlighted the main points in a quick and readable format, generally to be consumed within a minute. This two-step approach gave users flexibility of choosing between a quick overview of the policy event (displayed upfront while accessing the web application), and a more detailed story (displayed upon clicking a particular link). Similar to the L2 stories, the L1 stories also contained explanation of policy related jargon. Figure \ref{fig2} contains a snippet of an L1 story from the Defense theme of 2023. To check the quality of clusters formed, three annotators outside the author group manually checked 100 articles randomly selected from each cluster for belongingness. An inter-tagger agreement score of 0.89 was obtained, which reflects the strong consistency among annotators and the high quality of the resulting clusters.

\subsection{Numeric Summary} 
To enrich the user experience and provide deeper context, we implemented a numeric summary generation component. Using the L2 summaries, we extracted key numeric data points using a specialized prompt (see supplementary \cite{supplementary}), and presented in the form of key:value pairs like \textit{Defense Budget: INR 3.05 lakh crore}, allowing users to quickly grasp the numerical significance of policy decisions.  In future, we plan to enhance this feature by incorporating visual elements such as plots, charts, and interactive graphics to make the numerical information even more accessible and insightful for users.

\section{User Study}
\begin{figure*}[htbp]
    \centering
    
    \begin{minipage}{0.48\textwidth}
        \centering
        \begin{subfigure}{\textwidth}
            \includegraphics[width=\linewidth]{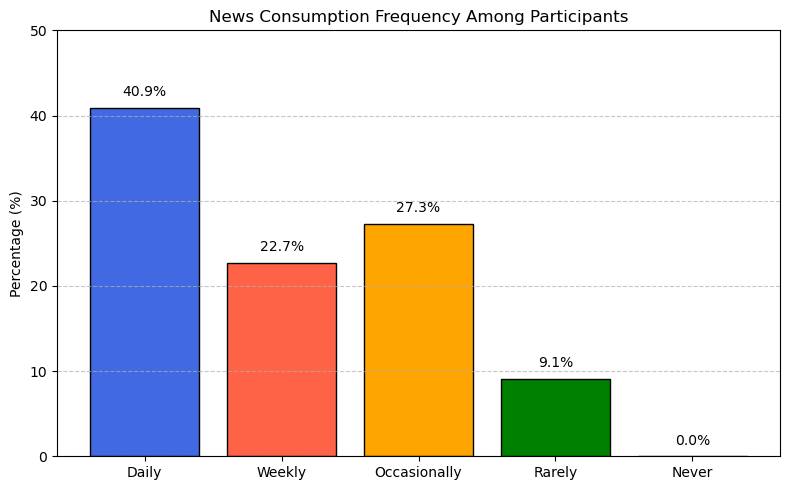}
            \caption{Frequency of news consumption.}
            \label{fig4}
        \end{subfigure}
        
        \vspace{0.5cm}
        
        \begin{subfigure}{\textwidth}
            \includegraphics[width=\linewidth]{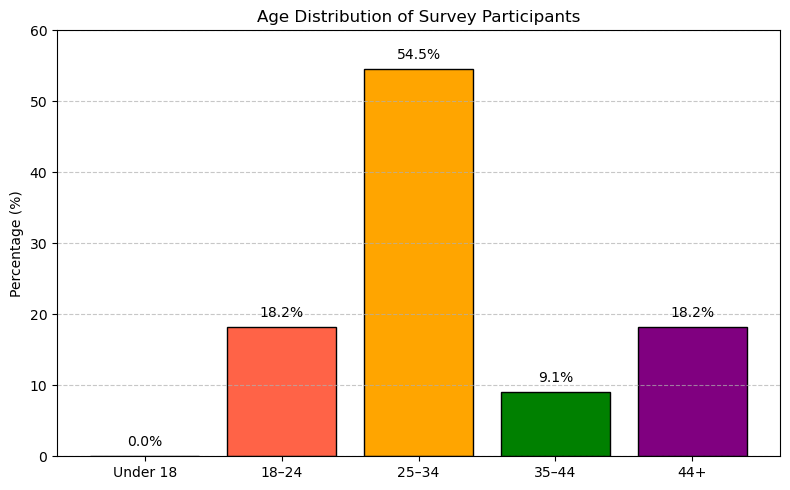}
            \caption{Age distribution of participants.}
            \label{fig5}
        \end{subfigure}
        
        \vspace{0.5cm}
        
        \begin{subfigure}{\textwidth}
            \includegraphics[width=\linewidth]{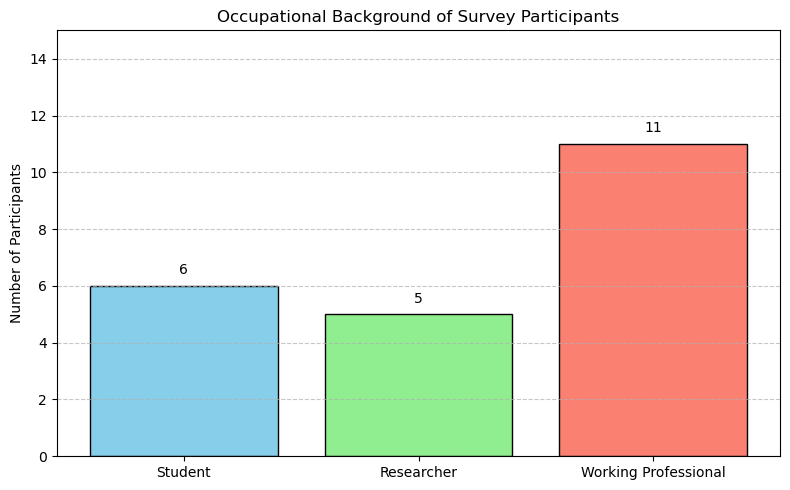}
            \caption{Participant professional affiliations.}
            \label{fig6}
        \end{subfigure}
        
        \label{fig:left_fig}
    \end{minipage}
    \hfill
    \begin{minipage}{0.48\textwidth}
        \centering
        \begin{subfigure}{\textwidth}
            \includegraphics[width=\linewidth]{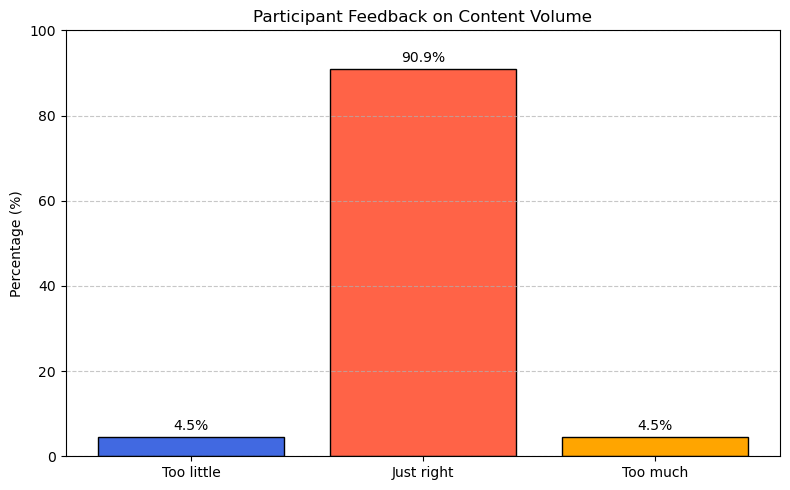}
            \caption{Feedback on content volume.}
            \label{fig7}
        \end{subfigure}
        
        \vspace{0.5cm}
        
        \begin{subfigure}{\textwidth}
            \includegraphics[width=\linewidth]{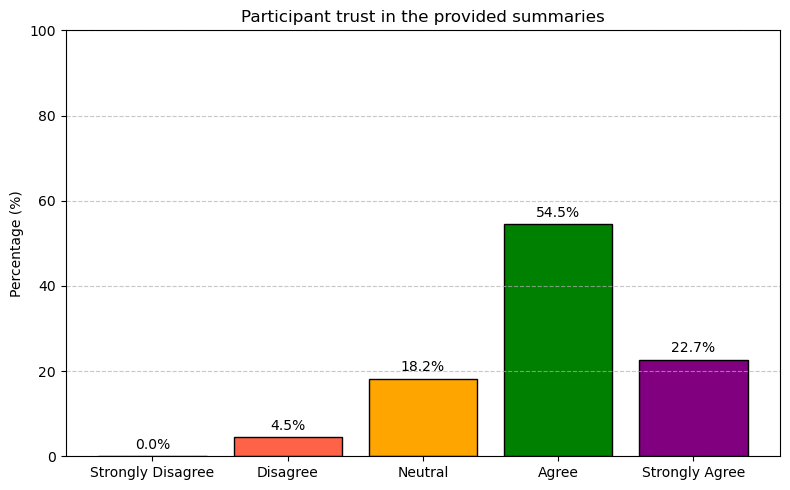}
            \caption{Perceived summary trustworthiness.}
            \label{fig8}
        \end{subfigure}
        
        \vspace{0.5cm}
        
        \begin{subfigure}{\textwidth}
            \includegraphics[width=\linewidth]{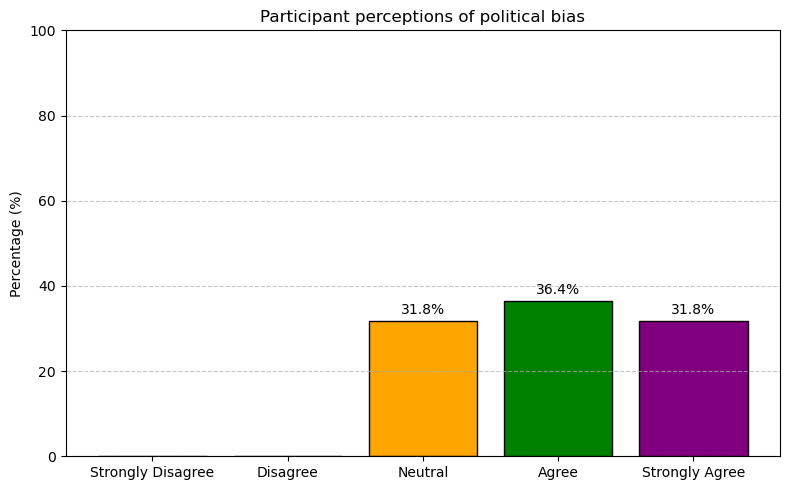}
            \caption{Perceptions of political bias.}
            \label{fig9}
        \end{subfigure}
        
        \label{fig:right_fig}
    \end{minipage}
\end{figure*}

We performed a user study around usability of PolicyStory, both in terms of the design of the tool and the comprehensibility, quality, and credibility of information presented. This was done using an online survey done within two academic institutes, and among people who generally knew about the policies studied. To avoid conflict of interests, we ensured that the participants were unrelated to the authors of the study, and were nudged anonymously for the survey. We also ensured that there was a healthy mix of participants who actively engaged regularly in news consumption on policies, and participants who seldom consumed news in general. As seen from Figure \ref{fig4}, 40.9\% of (9 of 22) participants in the survey were regular consumers of news (daily readers), while 27.3\% claimed to read news occasionally. This indicates a mix of informed and engaged user base, and casual consumers of news across platforms. Popular news sources included online print media, YouTube, and social media sources. In this section, we report the findings along different axes of evaluation of the tool. The detailed survey questions and responses are recorded in the supplementary material \cite{supplementary}.
\\
\\
\textbf{Demographics and News Consumption Habits:} Participants represented a diverse group, with the majority falling in the 25–34 age range (54.5\%, see Figure \ref{fig5}) and a mix of occupations including students, researchers, and working professionals (see Figure \ref{fig6}). The gender ratio among the participants was Male: 59.1\% and Female: 40.9\%. This ensured that the study spanned across multiple demographic groups, in terms of age, gender, and profession.
\\
\\
\textbf{Interface and Usability:} The interface was generally well-received, with 18 users rating it either 4 or 5 out of 5 for ease of use. Many praised its simplicity, noting that it helped focus attention on the summaries themselves. However, a few suggested small UX improvements, especially around the need for smoother transitions between summaries of different types:
\begin{quote}
\textit{P16: "Numeric summaries can be loaded on the same page, instead of loading a new page each time. same with the case of L2 summaries. UI can be more smooth in terms of loading L1 summaries. All the other details were fine. The colour gradient is a bit dull for the background, either keep it white or something which cause distraction."}
\end{quote}
\begin{quote}
\textit{P5: "That I have to choose a subtopic to move forward was not mentioned anywhere. I chose a topic from the dropdown menu and was waiting for the page to load on to the given topic."} 
\end{quote}
Another user pointed out the need to sort the L1 summaries in reverse chronological order, to prioritize recentness of information displayed.
\begin{quote}
    \textit{P20: "It might be better to have more recent stories towards the top since I was a bit confused as to why there was news about 2020 budget when I first saw the page and only got the full picture after scrolling down."}
\end{quote}
The issue reported by P5 is a genuine usability issue. However, we kept the interface as-is in this version, since displaying the summaries by default for a topic might also distract the user from exploring the other topics within a policy event. Loading each topic's details only upon clicking the topic button ensures uniformity in presentation. The chronological order of stories displayed (P20) was also deliberately retained, since the intention was to inform the user about the evolution of the issue over time. We do however intend to incorporate these feedback post completion of the broader user study, if more users report similarly. 
\\
\\
\textbf{L1 Summaries (Clarity and Impact):} The L1 summaries stood out for their clarity and time-saving nature. Nearly all users agreed that they provided a helpful snapshot of policy stories, with the important points highlighted as can be seen from the following quote.
\begin{quote}
\textit{P20: "... I think they had enough detail. The bullet points in the budget section also helped."}    
\end{quote}
Most participants agreed on the level of detail provided in L1 summaries to be perfect (see  Figure \ref{fig7}). It is also interesting to note that even from the brief L1 summaries, 95.5\% of participants claimed to have learnt something new about the policy issue. This suggests that the tool might aid in alleviating the problem of information overload, and actually inform the users new aspects about the policy issues that previously remained unexplored. Suggestions included making impacts of policy issues more explicit in the summary itself — one user noted:
\begin{quote}
    \textit{P2: "Impact and outcome of any policy should be  highlighted."}
\end{quote}
This feedback is indicative of the willingness of some participants (regular news consumers) to connect the news summaries to the actual impact of policy decisions. Performing information retrieval of official policy documents released by the Government and coupling the findings with the current summaries might help in this respect.
\\
\\
\textbf{L2 Summaries (Depth and Perspective):} Users who delved into the longer L2 summaries appreciated the structured detail. Similar to L1 summaries, almost all users agreed on the facts that the summaries provided them a clear overview of the policy, that they were interesting to read, and that they learnt something new about the policy. L2 summaries were deliberately presented in a story-like fashion to make the perusal interesting, which we believe resulted in user interest in the summaries, despite their length and level of detail. However, some users provided feedback on the flexibility aspect of L2 summaries as can be seen in the quote below.
\begin{quote}
    \textit{P5: "I think it can be a little more lengthy. I would like to suggest that L2 summaries can have 2-3 variations depending on reading time and readers can choose according to their time availability."}
\end{quote}
P5 being a regular news consumer, this feedback highlights the need for expert users to consume policy related news at multiple levels of depth. As part of future work, we plan to incorporate a user input of \textit{read-time}, which generates policy news summaries in real time based on user's bandwidth of information consumption. Another participant pointed out the need to highlight expert comments within the L2 summaries to ensure the richness of content presented (for example, comments by economists around the Union Budget).
\begin{quote}
    \textit{P2: "Experts' comment on the policy should be mentioned in the detailed summary."}
\end{quote}
Another participant (P6) mentioned that infographics can aid in making the L2 summaries more comprehensible. 
\\
\\
\textbf{Numeric Summaries:} While almost all users found the numeric summaries informative, providing them new information and a brief overview about the policy issues, a few others felt they lacked context (P5). This was expected since the numeric summaries were merely extracted from the L2 summaries and presented in the form of key:value pairs. For example, in the year 2023, expenditure around the Defense Budget was presented as \textit{Defense Budget: INR 5.94 lakh crore (INR 5.94 trillion)}. We plan to enrich the numeric summary with quantitative analysis and insights around key quantities with numeric estimates in the future version of the tool. Another user was unable to locate the numeric summary (P14), highlighting the need for stronger visual or contextual cues to highlight the utility of numbers. 
\\
\\
\textbf{Trust and Perception of Bias:} A majority felt the system was unbiased (see Figure \ref{fig9}), with 77.2\% of users reporting that the content felt trustworthy (see Figure \ref{fig8}). However, subtle tone issues did raise flags for a few (P5, P15). 
\begin{quote}
    \textit{P5: "I don't know if it was the selected topic because I read about union budget but it sounded a little anti-government. May be it was the language because we are used to more bland language in news."}
\end{quote}
P15 commented on how some of the summaries seemed more pro-farmers, in context of the Farmers' Protests. Although PolicyStory collects news from mainstream media sources, these perceptions underline the delicate role of tone and word choice in news summarization tools. Some participants (P14, P15) highlighted the need for citing appropriate sources in the summaries, indicative of their concerns around trustworthiness of summaries. This is also a feature that can be incorporated in the tool, since LLM based summarization can also aid in bringing out relevant sources and context for better reliance on the information presented.
\\
\\
\textbf{Overall Preference and Adoption:} We found that the reception of PolicyStory among the 22 participants surveyed was overwhelmingly positive. Almost all participants preferred using the tool regularly, over painstakingly consuming news from multiple news sources. 81.8\% users stated they would use the system regularly, and many (86.4\%) were enthusiastic about recommending it to others. This is an artifact of the information overload and complexity inherent in policy news consumption that PolicyStory attempts to alleviate -- the tool provided a concise, yet comprehensible, appropriate, and interesting overview of policy issues to the users, highlighting the role LLMs can play in making information easily consumable across user bases. We intend to incorporate some of the feedback around the usability and trustworthiness of the system in future versions of the system.

\section{Discussion and Conclusion}
In this paper, we proposed \textit{PolicyStory}, a system that summarizes Indian policy-related news using data from \textit{Media Cloud}, leveraging open-source LLMs (\textit{Llama-3.2-1B}). The tool presents a topic-wise, chronological, and longitudinal summary of policy discourse through a web interface. A preliminary user study indicated strong user preference for PolicyStory over reading multiple lengthy and jargon-heavy articles, particularly due to its structured and comprehensible presentation of evolving policy narratives.

PolicyStory offers three levels of summarization: a concise L1 summary, a detailed L2 (story-like) summary, and a numeric summary (key:value format). This design caters to varied user preferences in terms of depth. The tool addresses key challenges in policy news consumption—such as complexity, jargon, and temporal fragmentation—by providing clarity, contextual continuity, and explanation of terms. As policy understanding is vital to democratic participation, PolicyStory bridges the gap between complex media narratives and citizen comprehension of policy issues.

Compared to existing news summarization tools like \textit{Inshorts}\cite{reddy2024smart}, PolicyStory’s emphasis lies on long-term policy tracking rather than short-term event coverage. A noted concern is the credibility of sources within \textit{Media Cloud}, which includes both mainstream and smaller outlets. Future work includes integrating fact-checking modules to address this limitation~\cite{bib8,bib12,bib13}.

To enhance utility, incorporating expert commentary (e.g., from editorial pieces or policy documents) could help assess both media discourse and real-world impact. Interactive visualizations based on numeric insights, such as time trends in defense spending or protest participation, are also planned, leveraging LLMs to extract key quantitative insights.

Given the diverse user preferences and decreasing attention spans, we intend to incorporate a ``time-to-read” input feature in the tool. This would adjust the summary depth dynamically. While this introduces a trade-off between completeness and brevity, it aligns with real-world usage needs and warrants further exploration.

The user study, though limited in scale, confirmed the tool’s usefulness among laypersons and experts alike. We plan to expand the study across demographics. To evaluate information fidelity, we will adopt formal metrics like BERTScore, ROUGE, BLEURT, and coverage-based measures. Further, comparing PolicyStory’s readability with established readability scores~\cite{bib9,bib10} can validate its effectiveness in simplifying complex content.

Finally, we aim to extend PolicyStory’s coverage to include regional languages to better serve India’s diverse media audience and enhance the framework’s inclusivity and generalizability.

\backmatter

\bmhead{Supplementary information}

Additional details on the LLM prompts used, definition of topic clusters, web application, and user study can be found in the \href{https://drive.google.com/file/d/1frihWhA9thS3ntUMhuODb6TJbKKSyt68/view?usp=sharing}{supplementary material} \cite{supplementary}.

\bibliography{sn-bibliography}


\begin{thebibliography}{35}
\ifx \bisbn   \undefined \def \bisbn  #1{ISBN #1}\fi
\ifx \binits  \undefined \def \binits#1{#1}\fi
\ifx \bauthor  \undefined \def \bauthor#1{#1}\fi
\ifx \batitle  \undefined \def \batitle#1{#1}\fi
\ifx \bjtitle  \undefined \def \bjtitle#1{#1}\fi
\ifx \bvolume  \undefined \def \bvolume#1{\textbf{#1}}\fi
\ifx \byear  \undefined \def \byear#1{#1}\fi
\ifx \bissue  \undefined \def \bissue#1{#1}\fi
\ifx \bfpage  \undefined \def \bfpage#1{#1}\fi
\ifx \blpage  \undefined \def \blpage #1{#1}\fi
\ifx \burl  \undefined \def \burl#1{\textsf{#1}}\fi
\ifx \doiurl  \undefined \def \doiurl#1{\url{https://doi.org/#1}}\fi
\ifx \betal  \undefined \def \betal{\textit{et al.}}\fi
\ifx \binstitute  \undefined \def \binstitute#1{#1}\fi
\ifx \binstitutionaled  \undefined \def \binstitutionaled#1{#1}\fi
\ifx \bctitle  \undefined \def \bctitle#1{#1}\fi
\ifx \beditor  \undefined \def \beditor#1{#1}\fi
\ifx \bpublisher  \undefined \def \bpublisher#1{#1}\fi
\ifx \bbtitle  \undefined \def \bbtitle#1{#1}\fi
\ifx \bedition  \undefined \def \bedition#1{#1}\fi
\ifx \bseriesno  \undefined \def \bseriesno#1{#1}\fi
\ifx \blocation  \undefined \def \blocation#1{#1}\fi
\ifx \bsertitle  \undefined \def \bsertitle#1{#1}\fi
\ifx \bsnm \undefined \def \bsnm#1{#1}\fi
\ifx \bsuffix \undefined \def \bsuffix#1{#1}\fi
\ifx \bparticle \undefined \def \bparticle#1{#1}\fi
\ifx \barticle \undefined \def \barticle#1{#1}\fi
\bibcommenthead
\ifx \bconfdate \undefined \def \bconfdate #1{#1}\fi
\ifx \botherref \undefined \def \botherref #1{#1}\fi
\ifx \url \undefined \def \url#1{\textsf{#1}}\fi
\ifx \bchapter \undefined \def \bchapter#1{#1}\fi
\ifx \bbook \undefined \def \bbook#1{#1}\fi
\ifx \bcomment \undefined \def \bcomment#1{#1}\fi
\ifx \oauthor \undefined \def \oauthor#1{#1}\fi
\ifx \citeauthoryear \undefined \def \citeauthoryear#1{#1}\fi
\ifx \endbibitem  \undefined \def \endbibitem {}\fi
\ifx \bconflocation  \undefined \def \bconflocation#1{#1}\fi
\ifx \arxivurl  \undefined \def \arxivurl#1{\textsf{#1}}\fi
\csname PreBibitemsHook\endcsname

\bibitem[\protect\citeauthoryear{Das and Mishra}{2024}]{bib6}
\begin{botherref}
\oauthor{\bsnm{Das}, \binits{A.}},
\oauthor{\bsnm{Mishra}, \binits{M.H.}}:
The effect of reels on attention among young and middle-aged adults.
International Journal of Indian Psychology
\textbf{12}(3)
(2024)
\end{botherref}
\endbibitem

\bibitem[\protect\citeauthoryear{Mark}{2023}]{bib7}
\begin{bbook}
\bauthor{\bsnm{Mark}, \binits{G.}}:
\bbtitle{Attention Span: A Groundbreaking Way to Restore Balance, Happiness and Productivity}.
\bpublisher{Harlequin}, \blocation{???}
(\byear{2023})
\end{bbook}
\endbibitem

\bibitem[\protect\citeauthoryear{Singh}{2022}]{bib2}
\begin{barticle}
\bauthor{\bsnm{Singh}, \binits{S.}}:
\batitle{The farmers’ movement against three agricultural laws in india: A study of organisation}.
\bjtitle{Review of Agrarian Studies}
\bvolume{12}(\bissue{1}),
\bfpage{161}--\blpage{173}
(\byear{2022})
\end{barticle}
\endbibitem

\bibitem[\protect\citeauthoryear{Roberts et~al.}{2021}]{bib11}
\begin{bchapter}
\bauthor{\bsnm{Roberts}, \binits{H.}},
\bauthor{\bsnm{Bhargava}, \binits{R.}},
\bauthor{\bsnm{Valiukas}, \binits{L.}},
\bauthor{\bsnm{Jen}, \binits{D.}},
\bauthor{\bsnm{Malik}, \binits{M.M.}},
\bauthor{\bsnm{Bishop}, \binits{C.S.}},
\bauthor{\bsnm{Ndulue}, \binits{E.B.}},
\bauthor{\bsnm{Dave}, \binits{A.}},
\bauthor{\bsnm{Clark}, \binits{J.}},
\bauthor{\bsnm{Etling}, \binits{B.}}, \betal:
\bctitle{Media cloud: Massive open source collection of global news on the open web}.
In: \bbtitle{Proceedings of the International AAAI Conference on Web and Social Media},
vol. \bseriesno{15},
pp. \bfpage{1034}--\blpage{1045}
(\byear{2021})
\end{bchapter}
\endbibitem

\bibitem[\protect\citeauthoryear{Dar et~al.}{}]{PolicyStory}
\begin{botherref}
\oauthor{\bsnm{Dar}, \binits{A.N.}},
\oauthor{\bsnm{Singh}, \binits{A.}},
\oauthor{\bsnm{Sen}, \binits{A.}}:
{PolicyStory}.
\url{https://neural-times.vercel.app/}.
Online; accessed 11 June 2025
\end{botherref}
\endbibitem

\bibitem[\protect\citeauthoryear{Park et~al.}{2009}]{park2009newscube}
\begin{bchapter}
\bauthor{\bsnm{Park}, \binits{S.}},
\bauthor{\bsnm{Kang}, \binits{S.}},
\bauthor{\bsnm{Chung}, \binits{S.}},
\bauthor{\bsnm{Song}, \binits{J.}}:
\bctitle{Newscube: delivering multiple aspects of news to mitigate media bias}.
In: \bbtitle{Proceedings of the SIGCHI Conference on Human Factors in Computing Systems},
pp. \bfpage{443}--\blpage{452}
(\byear{2009})
\end{bchapter}
\endbibitem

\bibitem[\protect\citeauthoryear{Munson et~al.}{2013}]{munson2013encouraging}
\begin{bchapter}
\bauthor{\bsnm{Munson}, \binits{S.}},
\bauthor{\bsnm{Lee}, \binits{S.}},
\bauthor{\bsnm{Resnick}, \binits{P.}}:
\bctitle{Encouraging reading of diverse political viewpoints with a browser widget}.
In: \bbtitle{Proceedings of the International AAAI Conference on Web and Social Media},
vol. \bseriesno{7},
pp. \bfpage{419}--\blpage{428}
(\byear{2013})
\end{bchapter}
\endbibitem

\bibitem[\protect\citeauthoryear{Park et~al.}{2010}]{park2010aspect}
\begin{bchapter}
\bauthor{\bsnm{Park}, \binits{S.}},
\bauthor{\bsnm{Lee}, \binits{S.}},
\bauthor{\bsnm{Song}, \binits{J.}}:
\bctitle{Aspect-level news browsing: Understanding news events from multiple viewpoints}.
In: \bbtitle{Proceedings of the 15th International Conference on Intelligent User Interfaces},
pp. \bfpage{41}--\blpage{50}
(\byear{2010})
\end{bchapter}
\endbibitem

\bibitem[\protect\citeauthoryear{Reddy}{2024}]{reddy2024smart}
\begin{bchapter}
\bauthor{\bsnm{Reddy}, \binits{M.}}:
\bctitle{Smart news for the smartphones in the era of data journalism}.
In: \bbtitle{Handbook of Digital Journalism: Perspectives from South Asia},
pp. \bfpage{341}--\blpage{349}.
\bpublisher{Springer}, \blocation{???}
(\byear{2024})
\end{bchapter}
\endbibitem

\bibitem[\protect\citeauthoryear{Chuang et~al.}{2024}]{chuang2024spec}
\begin{barticle}
\bauthor{\bsnm{Chuang}, \binits{Y.-N.}},
\bauthor{\bsnm{Tang}, \binits{R.}},
\bauthor{\bsnm{Jiang}, \binits{X.}},
\bauthor{\bsnm{Hu}, \binits{X.}}:
\batitle{Spec: a soft prompt-based calibration on performance variability of large language model in clinical notes summarization}.
\bjtitle{Journal of biomedical informatics}
\bvolume{151},
\bfpage{104606}
(\byear{2024})
\end{barticle}
\endbibitem

\bibitem[\protect\citeauthoryear{Hartl and Kruschwitz}{2022}]{hartl2022applying}
\begin{botherref}
\oauthor{\bsnm{Hartl}, \binits{P.}},
\oauthor{\bsnm{Kruschwitz}, \binits{U.}}:
Applying automatic text summarization for fake news detection.
arXiv preprint arXiv:2204.01841
(2022)
\end{botherref}
\endbibitem

\bibitem[\protect\citeauthoryear{Zogan et~al.}{2021}]{zogan2021depressionnet}
\begin{bchapter}
\bauthor{\bsnm{Zogan}, \binits{H.}},
\bauthor{\bsnm{Razzak}, \binits{I.}},
\bauthor{\bsnm{Jameel}, \binits{S.}},
\bauthor{\bsnm{Xu}, \binits{G.}}:
\bctitle{Depressionnet: learning multi-modalities with user post summarization for depression detection on social media}.
In: \bbtitle{Proceedings of the 44th International ACM SIGIR Conference on Research and Development in Information Retrieval},
pp. \bfpage{133}--\blpage{142}
(\byear{2021})
\end{bchapter}
\endbibitem

\bibitem[\protect\citeauthoryear{Ghadimi and Beigy}{2022}]{ghadimi2022hybrid}
\begin{barticle}
\bauthor{\bsnm{Ghadimi}, \binits{A.}},
\bauthor{\bsnm{Beigy}, \binits{H.}}:
\batitle{Hybrid multi-document summarization using pre-trained language models}.
\bjtitle{Expert Systems with Applications}
\bvolume{192},
\bfpage{116292}
(\byear{2022})
\end{barticle}
\endbibitem

\bibitem[\protect\citeauthoryear{Ding et~al.}{2024}]{10685184}
\begin{bchapter}
\bauthor{\bsnm{Ding}, \binits{J.}},
\bauthor{\bsnm{Nguyen}, \binits{H.}},
\bauthor{\bsnm{Chen}, \binits{H.}}:
\bctitle{Evaluation of question-answering based text summarization using llm invited paper}.
In: \bbtitle{2024 IEEE International Conference on Artificial Intelligence Testing (AITest)},
pp. \bfpage{142}--\blpage{149}
(\byear{2024}).
\doiurl{10.1109/AITest62860.2024.00025}
\end{bchapter}
\endbibitem

\bibitem[\protect\citeauthoryear{Fang et~al.}{2024}]{fang2024multi}
\begin{botherref}
\oauthor{\bsnm{Fang}, \binits{J.}},
\oauthor{\bsnm{Liu}, \binits{C.-T.}},
\oauthor{\bsnm{Kim}, \binits{J.}},
\oauthor{\bsnm{Bhedaru}, \binits{Y.}},
\oauthor{\bsnm{Liu}, \binits{E.}},
\oauthor{\bsnm{Singh}, \binits{N.}},
\oauthor{\bsnm{Lipka}, \binits{N.}},
\oauthor{\bsnm{Mathur}, \binits{P.}},
\oauthor{\bsnm{Ahmed}, \binits{N.K.}},
\oauthor{\bsnm{Dernoncourt}, \binits{F.}}, et al.:
Multi-llm text summarization.
arXiv preprint arXiv:2412.15487
(2024)
\end{botherref}
\endbibitem

\bibitem[\protect\citeauthoryear{Shah et~al.}{2025}]{shah2025topic}
\begin{botherref}
\oauthor{\bsnm{Shah}, \binits{S.}},
\oauthor{\bsnm{Chandrasekaran}, \binits{D.}},
\oauthor{\bsnm{Ryali}, \binits{S.}},
\oauthor{\bsnm{Venkatesh}, \binits{R.}}:
Topic driven text summarization with defragmentation using llms.
Authorea Preprints
(2025)
\end{botherref}
\endbibitem

\bibitem[\protect\citeauthoryear{Sahu and Laradji}{2024}]{sahu2024mixsumm}
\begin{botherref}
\oauthor{\bsnm{Sahu}, \binits{G.}},
\oauthor{\bsnm{Laradji}, \binits{I.H.}}:
Mixsumm: Topic-based data augmentation using llms for low-resource extractive text summarization.
arXiv preprint arXiv:2407.07341
(2024)
\end{botherref}
\endbibitem

\bibitem[\protect\citeauthoryear{Song et~al.}{2024}]{song2024finesure}
\begin{botherref}
\oauthor{\bsnm{Song}, \binits{H.}},
\oauthor{\bsnm{Su}, \binits{H.}},
\oauthor{\bsnm{Shalyminov}, \binits{I.}},
\oauthor{\bsnm{Cai}, \binits{J.}},
\oauthor{\bsnm{Mansour}, \binits{S.}}:
Finesure: Fine-grained summarization evaluation using llms.
arXiv preprint arXiv:2407.00908
(2024)
\end{botherref}
\endbibitem

\bibitem[\protect\citeauthoryear{Jiang et~al.}{2024}]{jiang2024empirical}
\begin{bchapter}
\bauthor{\bsnm{Jiang}, \binits{Z.}},
\bauthor{\bsnm{Yang}, \binits{J.}},
\bauthor{\bsnm{Rao}, \binits{D.}}:
\bctitle{An empirical study of leveraging plms and llms for long-text summarization}.
In: \bbtitle{Pacific Rim International Conference on Artificial Intelligence},
pp. \bfpage{424}--\blpage{435}
(\byear{2024}).
\bcomment{Springer}
\end{bchapter}
\endbibitem

\bibitem[\protect\citeauthoryear{Dalecki et~al.}{2009}]{dalecki2009news}
\begin{barticle}
\bauthor{\bsnm{Dalecki}, \binits{L.}},
\bauthor{\bsnm{Lasorsa}, \binits{D.L.}},
\bauthor{\bsnm{Lewis}, \binits{S.C.}}:
\batitle{The news readability problem}.
\bjtitle{Journalism Practice}
\bvolume{3}(\bissue{1}),
\bfpage{1}--\blpage{12}
(\byear{2009})
\end{barticle}
\endbibitem

\bibitem[\protect\citeauthoryear{Kleinnijenhuis}{1991}]{kleinnijenhuis1991newspaper}
\begin{barticle}
\bauthor{\bsnm{Kleinnijenhuis}, \binits{J.}}:
\batitle{Newspaper complexity and the knowledge gap}.
\bjtitle{European journal of communication}
\bvolume{6}(\bissue{4}),
\bfpage{499}--\blpage{522}
(\byear{1991})
\end{barticle}
\endbibitem

\bibitem[\protect\citeauthoryear{Danielson et~al.}{1992}]{danielson1992journalists}
\begin{barticle}
\bauthor{\bsnm{Danielson}, \binits{W.A.}},
\bauthor{\bsnm{Lasorsa}, \binits{D.L.}},
\bauthor{\bsnm{Im}, \binits{D.S.}}:
\batitle{Journalists and novelists: A study of diverging styles}.
\bjtitle{Journalism Quarterly}
\bvolume{69}(\bissue{2}),
\bfpage{436}--\blpage{446}
(\byear{1992})
\end{barticle}
\endbibitem

\bibitem[\protect\citeauthoryear{Patoko and Yazdanifard}{2014}]{Patoko2014TheIO}
\begin{barticle}
\bauthor{\bsnm{Patoko}, \binits{N.}},
\bauthor{\bsnm{Yazdanifard}, \binits{R.}}:
\batitle{The impact of using many jargon words, while communicating with the organization employees}.
\bjtitle{American Journal of Industrial and Business Management}
\bvolume{04},
\bfpage{567}--\blpage{572}
(\byear{2014})
\end{barticle}
\endbibitem

\bibitem[\protect\citeauthoryear{Khawaja et~al.}{2014}]{Khawaja2014MeasuringCL}
\begin{barticle}
\bauthor{\bsnm{Khawaja}, \binits{M.A.}},
\bauthor{\bsnm{Chen}, \binits{F.}},
\bauthor{\bsnm{Marcus}, \binits{N.}}:
\batitle{Measuring cognitive load using linguistic features: Implications for usability evaluation and adaptive interaction design}.
\bjtitle{International Journal of Human-Computer Interaction}
\bvolume{30},
\bfpage{343}--\blpage{368}
(\byear{2014})
\end{barticle}
\endbibitem

\bibitem[\protect\citeauthoryear{Fix and Fairbanks}{2020}]{fix2020effect}
\begin{barticle}
\bauthor{\bsnm{Fix}, \binits{M.P.}},
\bauthor{\bsnm{Fairbanks}, \binits{B.R.}}:
\batitle{The effect of opinion readability on the impact of us supreme court precedents in state high courts}.
\bjtitle{Social Science Quarterly}
\bvolume{101}(\bissue{2}),
\bfpage{811}--\blpage{824}
(\byear{2020})
\end{barticle}
\endbibitem

\bibitem[\protect\citeauthoryear{Roberts et~al.}{2021}]{bib1}
\begin{bchapter}
\bauthor{\bsnm{Roberts}, \binits{H.}},
\bauthor{\bsnm{Bhargava}, \binits{R.}},
\bauthor{\bsnm{Valiukas}, \binits{L.}},
\bauthor{\bsnm{Jen}, \binits{D.}},
\bauthor{\bsnm{Malik}, \binits{M.M.}},
\bauthor{\bsnm{Bishop}, \binits{C.S.}},
\bauthor{\bsnm{Ndulue}, \binits{E.B.}},
\bauthor{\bsnm{Dave}, \binits{A.}},
\bauthor{\bsnm{Clark}, \binits{J.}},
\bauthor{\bsnm{Etling}, \binits{B.}}, \betal:
\bctitle{Media cloud: Massive open source collection of global news on the open web}.
In: \bbtitle{Proceedings of the International AAAI Conference on Web and Social Media},
vol. \bseriesno{15},
pp. \bfpage{1034}--\blpage{1045}
(\byear{2021})
\end{bchapter}
\endbibitem

\bibitem[\protect\citeauthoryear{Bhardwaj and Jadhav}{2024}]{bib3}
\begin{botherref}
\oauthor{\bsnm{Bhardwaj}, \binits{M.}},
\oauthor{\bsnm{Jadhav}, \binits{R.}}:
Why are indian farmers protesting again? demands for government explained.
Reuters
(2024).
Accessed: 2025-04-29
\end{botherref}
\endbibitem

\bibitem[\protect\citeauthoryear{Wolf et~al.}{2019}]{bib4}
\begin{botherref}
\oauthor{\bsnm{Wolf}, \binits{T.}},
\oauthor{\bsnm{Debut}, \binits{L.}},
\oauthor{\bsnm{Sanh}, \binits{V.}},
\oauthor{\bsnm{Chaumond}, \binits{J.}},
\oauthor{\bsnm{Delangue}, \binits{C.}},
\oauthor{\bsnm{Moi}, \binits{A.}},
\oauthor{\bsnm{Cistac}, \binits{P.}},
\oauthor{\bsnm{Rault}, \binits{T.}},
\oauthor{\bsnm{Louf}, \binits{R.}},
\oauthor{\bsnm{Funtowicz}, \binits{M.}}, et al.:
Huggingface's transformers: State-of-the-art natural language processing.
arXiv preprint arXiv:1910.03771
(2019)
\end{botherref}
\endbibitem

\bibitem[\protect\citeauthoryear{Grattafiori et~al.}{2024}]{bib5}
\begin{botherref}
\oauthor{\bsnm{Grattafiori}, \binits{A.}},
\oauthor{\bsnm{Dubey}, \binits{A.}},
\oauthor{\bsnm{Jauhri}, \binits{A.}},
\oauthor{\bsnm{Pandey}, \binits{A.}},
\oauthor{\bsnm{Kadian}, \binits{A.}},
\oauthor{\bsnm{Al-Dahle}, \binits{A.}},
\oauthor{\bsnm{Letman}, \binits{A.}},
\oauthor{\bsnm{Mathur}, \binits{A.}},
\oauthor{\bsnm{Schelten}, \binits{A.}},
\oauthor{\bsnm{Vaughan}, \binits{A.}}, et al.:
The llama 3 herd of models.
arXiv preprint arXiv:2407.21783
(2024)
\end{botherref}
\endbibitem

\bibitem[\protect\citeauthoryear{Dar et~al.}{}]{supplementary}
\begin{botherref}
\oauthor{\bsnm{Dar}, \binits{A.N.}},
\oauthor{\bsnm{Singh}, \binits{A.}},
\oauthor{\bsnm{Sen}, \binits{A.}}:
{Supplementary Material}.
\url{https://drive.google.com/file/d/1frihWhA9thS3ntUMhuODb6TJbKKSyt68/view?usp=sharing}.
Online; accessed 11 June 2025
\end{botherref}
\endbibitem

\bibitem[\protect\citeauthoryear{Diel and Roberts}{2021}]{bib8}
\begin{barticle}
\bauthor{\bsnm{Diel}, \binits{S.}},
\bauthor{\bsnm{Roberts}, \binits{C.}}:
\batitle{News story aggregation and perceived credibility}.
\bjtitle{Newspaper Research Journal}
\bvolume{42}(\bissue{2}),
\bfpage{162}--\blpage{181}
(\byear{2021})
\end{barticle}
\endbibitem

\bibitem[\protect\citeauthoryear{Huang et~al.}{2024}]{bib12}
\begin{barticle}
\bauthor{\bsnm{Huang}, \binits{H.}},
\bauthor{\bsnm{Zhu}, \binits{H.}},
\bauthor{\bsnm{Liu}, \binits{W.}},
\bauthor{\bsnm{Gao}, \binits{H.}},
\bauthor{\bsnm{Jin}, \binits{H.}},
\bauthor{\bsnm{Liu}, \binits{B.}}:
\batitle{Uncovering the essence of diverse media biases from the semantic embedding space}.
\bjtitle{Humanities and Social Sciences Communications}
\bvolume{11}(\bissue{1}),
\bfpage{1}--\blpage{12}
(\byear{2024})
\end{barticle}
\endbibitem

\bibitem[\protect\citeauthoryear{Shah et~al.}{2024}]{bib13}
\begin{bchapter}
\bauthor{\bsnm{Shah}, \binits{B.S.}},
\bauthor{\bsnm{Shah}, \binits{D.S.}},
\bauthor{\bsnm{Attar}, \binits{V.}}:
\bctitle{Decoding news bias: Multi bias detection in news articles}.
In: \bbtitle{Proceedings of the 2024 8th International Conference on Natural Language Processing and Information Retrieval},
pp. \bfpage{97}--\blpage{104}
(\byear{2024})
\end{bchapter}
\endbibitem

\bibitem[\protect\citeauthoryear{van Schaik and Pugh}{2024}]{bib9}
\begin{bchapter}
\bauthor{\bsnm{Schaik}, \binits{T.A.}},
\bauthor{\bsnm{Pugh}, \binits{B.}}:
\bctitle{A field guide to automatic evaluation of llm-generated summaries}.
In: \bbtitle{Proceedings of the 47th International ACM SIGIR Conference on Research and Development in Information Retrieval},
pp. \bfpage{2832}--\blpage{2836}
(\byear{2024})
\end{bchapter}
\endbibitem

\bibitem[\protect\citeauthoryear{Crossley et~al.}{2011}]{bib10}
\begin{barticle}
\bauthor{\bsnm{Crossley}, \binits{S.A.}},
\bauthor{\bsnm{Allen}, \binits{D.B.}},
\bauthor{\bsnm{McNamara}, \binits{D.S.}}:
\batitle{Text readability and intuitive simplification: A comparison of readability formulas.}
\bjtitle{Reading in a foreign language}
\bvolume{23}(\bissue{1}),
\bfpage{84}--\blpage{101}
(\byear{2011})
\end{barticle}
\endbibitem

\end{thebibliography}

\end{document}